\documentclass{elsart}

\journal{Physics Letters B}

\usepackage{amssymb}

\newcommand{\Gmatrixelement}{\bra{k p} G\left(e(h)+e(h')\right) \ket{h h'}}

\newcommand{\energydenominator}{E-e(p)+e(h)+e(h')}
\newcommand{\smallfrac}[2]{\textstyle{{\scriptstyle #1}\over {\scriptstyle#2}}}
\newcommand{\bra}[1]{\langle #1 \vert}
\newcommand{\ket}[1]{\vert #1 \rangle}
\newcommand{\beq}{\begin{equation}}
\newcommand{\eeq}{\end{equation}}
\def\bar{\begin{array}}
\def\ear{\end{array}}

\def\newline{\hfil\break}
\def\newpage{\vfill\eject}

\begin{document}

\begin{frontmatter}

\title{Nuclear matter hole spectral function in the Bethe-Brueckner-Goldstone
approach}

\author{M. Baldo\corauthref{cor}}
\address{INFN, Sezione di Catania, 57 Corso Italia, I-95129 Catania, Italy}
\corauth[cor]{Corresponding author - Fax: ++39 95 383023}
\ead{BALDO@CT.INFN.IT}

\author{L. Lo Monaco}
\address{Dipartimento di Fisica e Astronomia, Universit\`a di Catania, and 
INFN, Sezione di Catania, 57 Corso Italia, I-95129 Catania, Italy}
\ead{LOMONACO@CT.INFN.IT}

\begin{abstract}

The hole spectral function is calculated in nuclear matter to assess the
relevance of nucleon-nucleon short range correlations. The calculation is
carried out within the Brueckner scheme of many-body theory by using several
nucleon-nucleon realistic interactions. Results are compared with other
approaches based on variational methods and transport theory. Discrepancies
appear in the high energy region, which is sensitive to short range
correlations, and are due to the different many-body treatment more than to the
specific N-N interaction used. Another conclusion is that the momentum
dependence of the G-matrix should be taken into account in any self consistent
approach.

\end{abstract}

\end{frontmatter}

\section{Introduction}
\label{1}
One of the main features of nuclear systems is the relevance of short range
correlations (SRC) due to the hard core part of the nucleon-nucleon (NN)
potential. These are relevant for the binding energy of the nuclear system in
general and are probed directly by the high momentum and energy region of the
hole spectral function. In fact, SRC are entirely responsible for this region,
where the mean field and the long range correlations can give only negligeable
contributions. Spectral functions are extracted mainly from electron-nucleus
scattering experiments. \newline
Theoretically, many approaches have been proposed for a quantitative
description of SRC and the corresponding spectral functions. The many-body
aspect appears essential for a correct understanding of the correlations, and
Bethe-Brueckner \cite{Mahaux} or variational \cite{Fant1,Fant2} methods 
have been
extensively applied. Although the main features of the spectral functions in
the region relevant for SRC seem to have a similar trend within different
approaches, discrepancies are still present and controversial interpretations
of the experimental data exist in the literature. 
Recently \cite{Mosel} it was pointed out that the main features of the nucleon
spectral function in nuclear matter can be understood in the framework of
transport theory. In this approach the only ingredient is the in-medium
nucleon-nucoleon cross section, which can be, in first approximation,
schematized by a single parameter throughout the entire momentum and energy
range. The self-consistent calculation of the single particle self-energy and
strength function then provides the nucleon spectral function. Therefore, phase
space effects seem to dominate the behaviour of the spectral function.\par
A close comparison among the results of different approaches appears necessary
to extract the characteristics of SRC which determine the strength function.
In particular, the dependence of the results on the particular many-body theory
as well as on the specific nucleon-nucleon interaction used in the calculations
should elucidate the structure of the SRC.\par
In this letter, we present calculations of the nucleon spectral function in
nuclear matter within the Brueckner scheme using different realistic NN
interactions and compare them with other approaches. We will 
concentrate the comparison mainly in the
high energy region, where the effect of correlations is dominating.

\section{Formalism}
The hole spectral function in nuclear matter is defined as:
\begin{equation}
S(k,\omega) = \sum_f{ \left| \langle \Psi^f_{A-1} \vert a_k 
\vert \Psi^0_A \rangle \right| ^2 
\delta \left( \omega - \left( E^f_{A-1}-E^0_A \right) \right)}
\end{equation}
where $\ket{\Psi^0_A}$ is the ground state for $A$ nucleons
with eigenvalue $E^0_A$, $\ket{\Psi^f_{A-1}}$ is a complete set 
of states for $A-1$ nucleons with eigenvalues $E^f_{A-1}$, and
$a_k$ is the annihilation operator for the normalized plane
wave at momentum $k$.\par
In a single-particle description of nuclei only states with $k<k_F$ are 
occupied and the final states involved in the sum are at most 
one hole states. In a correlated nucleus, the wave function of the 
ground state can have high momentum components, and the structure of the 
spectral function becomes more complicated.
\par
In nuclear matter, the spectral function corresponding to the nucleon
self-energy $M(k,\omega) = V(k,\omega) + iW(k,\omega)$, 
is given by the well known result \cite{Mahaux}:
\begin{equation}
S(k,\omega) = - \frac{1}{\pi} {\rm Im} \, {\mathcal G}(k,\omega) = 
- \frac{1}{\pi} \frac{W(k,\omega)}{ (-\omega-\frac{k^2}{2m}-V(k,\omega))^2+
W(k,\omega)^2} 
\end{equation}
\noindent
where ${\mathcal G}(k,\omega)$ is the single--particle Green's function:
\begin{equation}
{\mathcal G}(k,\omega)=\frac{1}{-\omega-\frac{k^2}{2m} - V(k,\omega) - 
{\rm i} W(k,\omega)}
\end{equation}
\noindent
The real  and imaginary parts of the self-energy, $V(k,\omega)$ and 
$W(k,\omega)$, are highly off-shell in the considered energy and momentum 
ranges. 
In nuclear matter, according to the BBG expansion, the whole set of two-hole 
line contributions is summed up by the diagrams depicted in
Fig.~1\ref{fig1} \cite{Mahaux}, where  the wavy lines indicate the Brueckner
 G-matrix. 
Since it is known that correlations with higher number of holes give a much
smaller contribution to the ground state energy~\cite{Day2,bal1,bal2}, it is 
reasonable 
to assume that their effect on the spectral function is also of minor 
relevance, producing mainly an additional background. 
The first diagram $(1.a)$ corresponds to the standard Brueckner
approximation for the nucleon self-energy, and has the following
expression
\beq
M_1 (k,\omega)\, =\, \sum_{k' < k_F} \bra{kk'}  G (e(k')+\omega) 
\ket{kk'}_a
\eeq
where $e(k)$ is the self-consistent single particle energy and the label $a$
means antisymmetrization. In the energy range relevant to the hole strength
function and for $k > k_F$, this diagram contributes only to the real part
$V(k,\omega)$ of the self-energy. Indeed, the imaginary part of $M_1(k,\omega)$
comes only from on-shell transitions from one-particle states ($k > k_F$) to
two-particle one-hole states and this process is not allowed by momentum
conservation. The diagram $(1.b)$ takes into account transitions to one-particle
two-hole states. It contributes to both real and imaginary parts in the
considered energy and momentum range. Explicitely, the imaginary part
$W(k,\omega)$ is given by
\begin{eqnarray}
&W(k,\omega) = \half \sum_{h h' p} {\rm Im}
\frac
{\vert\Gmatrixelement_a\vert^2}
{\energydenominator - {\rm i} \eta } =\nonumber\\
&\smallfrac {\pi}{2} \sum_{h h' p} 
{\vert \Gmatrixelement_a \vert^2} \, \delta ({-\omega + e(p) - e(h) - e(h') })
\end{eqnarray}
\noindent
where the sum is restricted
to states $h$ and $h'$ with a momentum smaller than $k_F$ and $p$
with momentum larger than $k_F$.\par
Once the Brueckner calculation for nuclear matter is performed,
the G-matrix and the corresponding self-consistent single particle spectrum
are obtained. The G-matrix is then calculated off-shell for the entry
energies needed in Eqs. (4,5), which provide the real and imaginary parts
of the self-energy. Finally, the spectral function is calculated 
from Eq. (2).\par
Equation (5) is similar to the formula used in ref. \cite{Mosel}. 
The main difference is the inclusion in the latter of the single particle 
strength functions in the phase space integral, which implies a
self-consistent calculation. Furthermore, in that work the square of the 
in-medium scattering matrix is approximated by a constant average value,
which simplifies  the calculation considerably. The average scattering matrix
is then adjusted in order to reproduce at best the results 
of ref. \cite{Ciofi}. \par
Equation (5) looks also similar to the formula for perturbative corrections 
introduced in ref. \cite{Fant2} within the variational scheme. 
In this case, the g-matrix is replaced by the residual interaction between 
hole states and two-hole one-particle states. The perturbative corrections are
essential near the Fermi energy since they allow the hole states to
acquire a width. However, in the high energy region, where we are 
focusing the analysis, 
their contribution is only marginal. In the variational scheme the main 
contribution comes from the two-hole one-particle orhtogonal correlated 
intermediate states, which cannot be cast in a formula like Equation (5).
Therefore a formal comparison between the BBG and variational
approach would be misleading. \par
As a consequence the numerical comparison among these many-body 
theories appears necessary.

\section{Results}
We performed calculations of the nuclear matter spectral function with three
different two-body potentials, the Urbana v$_{14}$ \cite{urbana},
the Argonne v$_{14}$ \cite{arg14} and the Argonne v$_{18}$ \cite{arg18}. 
Three-body
forces were added, according to the Urbana IX model \cite{ficks}, and adjusted
to reproduce the correct saturation point. We checked that
the effect of the three-body force on the spectral function is negligeable.
The Urbana potential was used only to compare with the results of
ref. \cite{Fant2} and ref. \cite{Mosel}. \par
In Figs.~2-a and 2-b we show the spectral function calculated at two
values of the momentum, $k = 2.25$ fm$^{-1}$ and $k = 3.5$ fm$^{-1}$. The three lines
indicate our BBG calculations with different potentials as described in the
figure, while the full circles label the results of ref. \cite{Fant2}, where
the Urbana v$_{14}$ was used. For both momenta the dependence on the
nucleon-nucleon potential appears quite weak, only in some cases a discrepancy
is present, which however does not exceed 20\%. Larger deviations in the high
energy region occur between the BBG results and those of ref. \cite{Mosel}. We
can conclude that these discrepancies in the high energy region are due to the
many-body treatment and not to the interaction employed.
\par
In Figs.~3-a and 3-b we compare our results (for the Urbana v$_{14}$
potential) to the fully self consistent calculation of Lehr et al. \cite{Mosel}.
In this case the comparison 
can be more transparent. In fact, the first order calculations of ref. 
\cite{Mosel}, indicated by the dotted lines in the figures, are exactly 
equivalent to our calculations if the G-matrix in Eq. (5) is replaced by
a constant average value independent of momenta, directly related to in-medium
nucleon-nucleon cross section \cite{Mosel}. The discrepancy with the BBG 
calculations is, in this case, only due to this approximation, i.e. to the
neglect of the G-matrix momentum dependence. This shows the relevance of the
momentum dependent correlations for the determination of the single 
particle spectral function. The fact that the introduction of the 
self-consistency moves the high energy tail towards the BBG behaviour 
appears misleading, since in the BBG calculations no self-consistency has 
been used. This observation seems equally well applicable to the comparison 
with the variational calculations \cite{Fant2}, where also no
self-consistency  procedure is included in the many-body scheme.\par
Of course these conclusions do not imply that the self-consistent procedure
is not relevant, but only that it must be carried out with the full momentum
dependence of the G-matrix.

\vfill\eject

\begin{figure}
\leftline{\bf Figure Captions.}
\vskip 1 true cm
\caption{Diagrams contributing to the nucleon self-energy $M(k,\omega)
 = V(k,\omega) + iW(k,\omega)$. The wavy lines indicate the Brueckner G-matrix.
(a)\ \ The first diagram is the standard Brueckner approximation
$M_1(k,\omega)$ for the nucleon self-energy, eq.(4). This diagram contributes
only to the real part $V(k,\omega)$ of the self-energy.
(b)\ \ This diagram takes into account transitions to one-particle
two-hole states. It contributes to both real and imaginary parts of the
self-energy in the considered energy and momentum range.}
\label{fig1}

\caption{BBG calculations of the nuclear matter strength function $S(k,\omega)$.
The solid line is our result with the Argonne v$_{18}$ potential, the
long-dashed line is obtained with Argonne v$_{14}$ and the dot-dashed line with
the Urbana v${14}.$ The black dots are the variational results of Benhar et
al. 
(a)\ \ For the momentum $k = 2.25$ fm$^{-1}.$ 
(b)\ \ For $k = 3.5$ fm$^{-1}.$} 
\label{fig2}

\caption{Comparison between our results and those of Lehr et al., ref. [4]. 
The dotted curves are obtained in that work after the first iteration step,
the solid curves are their fully self consistent calculation; the dashed curves
show the variational results of Benhar et al. , ref. [3]. 
Our BBG results calculated with the Urbana potential are shown here as
asterisks. 
(a)\ \ For the momentum $k = 2.26$ fm$^{-1}.$ 
(b) For $k = 3.59$ fm$^{-1}.$ }
\label{fig3}
\end{figure}

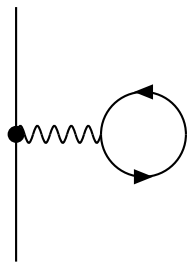
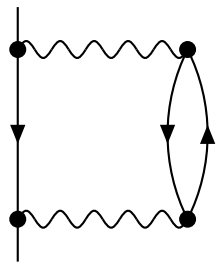
\end{document}